# A contact-waiting-time metric and RNA folding rates

Asamoah Nkwanta[1] and Wilfred Ndifon[2,*]

[1]Department of Mathematics, Morgan State University, Baltimore, MD 21251

[2]Department of Ecology & Evolutionary Biology, Princeton University, Princeton, NJ 08544

## Abstract

Metrics for indirectly predicting the folding rates of RNA sequences are of interest. In this letter, we introduce a simple metric of RNA structural complexity, which accounts for differences in the energetic contributions of RNA base contacts toward RNA structure formation. We apply the metric to RNA sequences whose folding rates were previously determined experimentally. We find that the metric has good correlation (correlation coefficient: -0.95, $p \ll 0.01$) with the logarithmically transformed folding rates of those RNA sequences. This suggests that the metric can be useful for predicting RNA folding rates. We use the metric to predict the folding rates of bacterial and eukaryotic group II introns. Future applications of the metric (e.g., to predict structural RNAs) could prove fruitful.

## INTRODUCTION

RNA molecules perform a variety of catalytic functions in living cells [1,2], mediated by well-defined structures. Knowledge of the native RNA secondary/tertiary structures and the rates at which RNA sequences fold into these structures is very useful for various practical applications and theoretical studies of RNA molecular biology and evolution. In general, the rate of folding into either the secondary or tertiary structure can be experimentally predicted for RNA sequences of varying lengths (e.g., see [3,4]). In contrast, it is very difficult to predict the folding rate for reasonably long RNA sequences by theoretical means. This is due, particularly, to the large number of very long, independent stochastic simulations required to predict the folding rates of such sequences [5]. In order to circumvent these computational difficulties, attempts have been made to develop indirect metrics for predicting RNA folding rates with fewer computational demands. One such metric [6] requires knowledge of both the collapse and folding transition temperatures of the RNA sequence whose folding rate is of interest. These temperatures are not always easy to estimate. Another metric is the relative contact order (CO), defined as:

$$\text{Relative CO} = \frac{1}{N_c \cdot L} \sum_{\text{contacts}} \Delta L_{ij} , \qquad (1)$$

where $N_c$ is the total number of contacts found in the structure of the RNA sequence under consideration, $L$ the length of the sequence, and $\Delta L_{ij}$ the number of bases found between the $i^{\text{th}}$ and $j^{\text{th}}$ contacting bases [3,7].

[*]To whom correspondence should be addressed. E-mail: ndifon@gmail.com

The relative CO was previously shown to correlate well with the logarithm of experimentally determined folding rates of short protein sequences [7]. A seminal study that applied this metric to 10 RNA sequences of varying lengths demonstrated, for four of the sequences, good correlation with the logarithm of the experimentally determined RNA folding rate [3]. The results of this particular study suggested that the RNA sequences could be divided into two classes, one consisting of six sequences that fold rapidly and in a manner that is independent of the relative CO, and the other consisting of four sequences that fold slowly, at rates that correlate with the relative CO. A similar classification of the sequences was suggested by results obtained using a variant of the relative CO, called the reduced CO. The reduced CO is also given by (1), with the important difference that only non-Watson-Crick contacts are considered. The motivation for the reduced CO is that non-Watson-Crick contacts may occur during tertiary (as opposed to secondary) structure formation. If tertiary structure formation constitutes the rate-limiting step in the RNA folding process, then consideration of non-Watson-Crick contacts may improve the ability to predict RNA folding rates [3].

However, previous work (e.g., [8,9]) suggests that there can be considerable RNA secondary structure rearrangement following the formation of tertiary contacts, and mutations that stabilize the native secondary structure can substantially increase the overall RNA folding rate, suggesting that tertiary structure formation is not necessarily folding-rate-limiting. In this letter, we show that by accounting for differential contributions of both Watson-Crick and non-Watson-Crick contacts toward the stabilization of RNA structures the folding rates of the RNA sequences analyzed in [3] can be predicted with reasonable accuracy.

## RESULTS AND DISCUSSION

Both the relative and reduced CO metrics described above do not account for the differential energetic contributions of RNA base contacts, and for the entropic costs associated the nucleation of RNA helices. These parameters are critical to RNA folding kinetics [10]. In particular, RNA folding involves the formation and dissociation of base contacts at rates that depend on the contacting bases, the structural context, and the folding environment (e.g., temperature and ionic concentration). The formation of isolated base pairs, which have the potential to nucleate new RNA helices, is associated with a loss of RNA conformational entropy and is therefore unfavorable. This entropic cost of helix nucleation can be offset to some degree by favorable energetic contributions, resulting from base stacking interactions the magnitudes of which depend on the bases involved [10]. Therefore, accounting for the different energetic contributions of base contacts as well as the entropic costs associated with helix nucleation can lead to improved prediction of RNA folding rates.

Let $\Delta G_{ij}$ denote the energetic contributions due to the $i^{\text{th}}$ and $j^{\text{th}}$ contacting bases. The rate of formation of this contact can be approximated by $\exp\left(-\Delta G_{ij}/(RT)\right)$, where $R$ denotes the gas constant and $T$ denotes the absolute temperature. Hence, the expected waiting time until the given contact is formed is $\sim\exp\left(\Delta G_{ij}/(RT)\right)$. The total waiting time until the formation of all contacts can be approximated by the sum of the waiting times associated with each contact[a]. For a helix-nucleating contact, which is presumably the contact that is separated by the smallest number of intervening bases [10], the waiting time may increase with the distance between the contacting bases. We account for this fact by weighting the energetic contributions resulting from such a nucleating contact between bases $i$ and $j$ by the logarithm of the distance $d_{ij}$ between the bases[b].

---

[a] This assumes that base contacts do not all form at once.

[b] In practice, the type of transformation that is applied to the distance should be dictated by the magnitude of the energetic contributions associated with base contacts; the magnitude of the distance should not be substantially greater than the magnitude of the energetic contributions.

More specifically, we approximate the folding time (i.e., the reciprocal of the folding rate) of an RNA sequence by the following contact-waiting-time (CWT) metric:

$$\mathrm{CWT} = \sum_{\text{contacts}} \exp\left(\Delta G_{ij} / (RT)\right), \qquad (2)$$

$$\Delta G_{ij} = \begin{cases} \sigma_{ij} / d_{ij}, \text{ for helix - nucleating contacts} \\ \sigma_{ij}, \text{ for non - nucleating contacts} \end{cases} \qquad (3)$$

We employ one of the simplest biophysically motivated[c] assignments of energetic contributions to base contacts: $\sigma_{ij}$ equals -1, -2, and -3kcal/mol for GU, AU, and GC contacts, respectively. In addition, we set $\sigma_{ij} = 1$, for mismatched contacts between identical bases, and we do not assign energetic contributions to other types of contacts. We use $T$ = 37$^{o}$C = 310K. A MATLAB code that calculates the CWT given an RNA sequence and secondary structure accompanies this letter as supplementary material.

We applied the above CWT metric to the RNA sequences previously analyzed in [3] with the goal of predicting the logarithm of the folding rates of those sequences. The results (see Table 1 & Figure 1) show good correlation (correlation coefficient: -0.95, $p$ << 0.01) between the CWT and the logarithm of the folding rate. This was better than the correlation (correlation coefficient: -0.39) obtained by using the reduced CO (see Figure 1). These results suggest that the incorporation of (sequence-based) information about differences in the energetic contributions of base contacts into measures of RNA structural complexity, such as the relative and reduced CO metrics [3,7], could perhaps improve the prediction of RNA folding rates with these two metrics. The usefulness of such improved metrics for the prediction of, for example, genomic sequences that encode functional RNA molecules and the elucidation of biophysical constraints on RNA evolution are interesting topics for future research.

[c] GU, AU, and GC contacts involve the formation of one, two, and three hydrogen bonds, respectively, and each hydrogen bond can contribute up to 1kcal/mol to the thermal stability of an RNA structure [10].

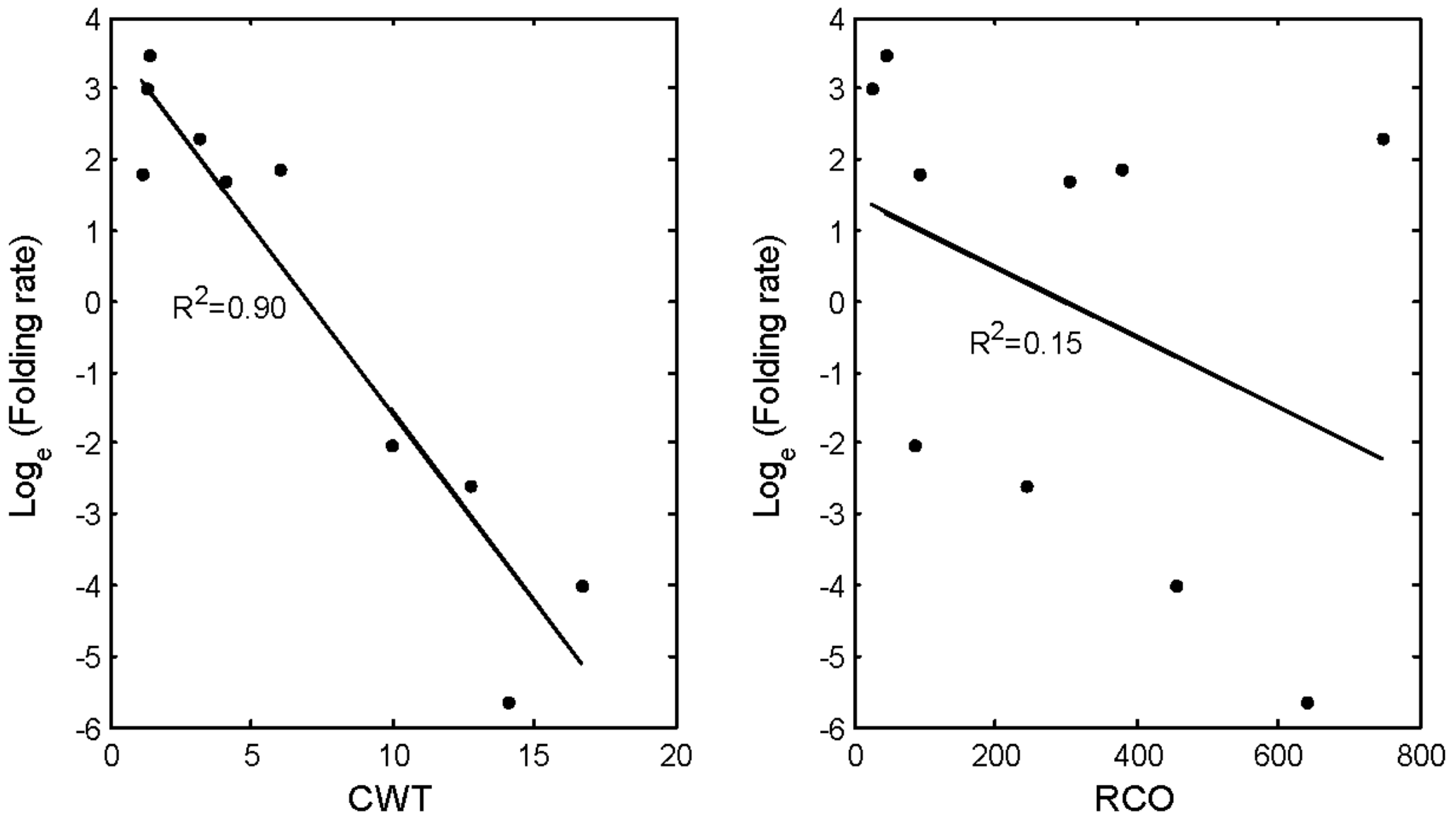

Figure 1. Relationship between the logarithm (to base e) of the RNA folding rate (in units of $sec^{-1}$) and both the contact-waiting-time (CWT) metric and the reduced contact order (RCO). Empirical estimates of the folding rate and the CWT are given in Table 1. Estimates of the RCO were extracted from Figure 3 of [3]. The figure shows a strong linear relationship between the logarithm (base e) of the folding rate and the CWT: $\log_e$(folding rate) = $-5.2798\times10^{-1}$ CWT + 3.7118. $R^2$ denotes the square of the correlation coefficient.

Table 1. Contact waiting time (CWT) and folding rates of RNA sequences

| **Name of RNA** | **CWT** | **Log$_e$ (Folding rate in units of $sec^{-1}$)** |
|---|---|---|
| *B. Subtilis* P RNA | 14.109 | -5.655 |
| *B. Subtilis* specificity-domain | 12.747 | -2.590 |
| *B. Subtilis* C-domain | 6.008 | 1.872 |
| *Azoarcus* group I intron | 3.147 | 2.303 |
| Hairpin ribozyme, four-way junction | 1.142 | 1.792 |
| *Tetrahymena* group I intron P5abc domain | 1.424 | 3.466 |
| *Tetrahymena* group I intron P4-P6 domain | 4.102 | 1.705 |
| *S. cereviseae* phenylalanil-tRNA | 1.338 | 2.996 |
| Hairpin ribozyme, two-way junction | 16.687 | -4.017 |
| *Tetrahymena* group I intron | 9.921 | -2.040 |

CWT was computed using Eqn. (2). Experimentally determined folding rates were copied without modification from [3], except in the case of the hairpin ribozyme, two-way junction whose folding rate was calculated as min($K_{dock}$, $K_{undock}$) rather than ($K_{dock}+K_{undock}$), as was done in [3] (Note that the minimum of $K_{dock}$ and $K_{undock}$ is rate limiting. Hence, it is a more appropriate estimate of the folding rate). The sequences and secondary structures of the displayed RNAs are given in Table S1.

To illustrate one possible application of the CWT metric, we used the above mentioned linear relationship between the metric and the folding rate (see the legend of Figure 1) to predict the folding rates of group II introns from a variety of species. Group II introns are large RNA molecules that perform a variety of catalytic functions in bacteria, lower eukaryotes, and plants [11]. Existing estimates of the folding rates of these important RNAs are largely based on the ai5γ model group II intron derived from *Saccharomyces cerevisiae* [11]. This model intron, which contains only a subset of the six domains normally found in the structure of the intact intron, was found to fold slowly.

Knowledge of the folding rates of intact group II introns could be useful. Therefore, we downloaded the sequences and secondary structures of group II introns from the Comparative RNA database of Gutell and co-workers [12]. We predicted the folding rate (at 37$^{o}$C) of the first intron listed for each species (see Table 2). The results suggest that all the analyzed introns fold slowly relative to other known RNAs. The harmonic mean of the predicted folding rates is ~$2.4\times10^{-3}$sec$^{-1}$, which is within an order of magnitude of the folding rate of ai5γ (i.e., ~$1.7\times10^{-2}$sec$^{-1}$), determined experimentally at 42$^{o}$C [11]. These predictions can be tested experimentally.

Table 2. Predicted folding rates of group II introns from different species

| **Name of RNA (length)** | **Species** | **CWT** | **Predicted folding rate (sec$^{-1}$)** |
|---|---|---|---|
| a.I2.b.C.sp.B.TBD.i5 (2543) | *Calothrix sp.* | 13.461 | 3.353e-2 |
| a.I2.b.E.coli.A.TBD.i1 (1979) | *Escherichia coli* | 16.677 | 6.137e-3 |
| a.I2.b.L.lactis.A.LtrB.i1 (2600) | *Lactococcus lactis* | 15.860 | 9.448e-3 |
| a.I2.c.N.tabacum.A.A6.i1 (780) | *Nicotiana tabacum* | 12.424 | 5.797e-2 |
| a.I2.m.A.aegerita.B.LSU.2059.bpseq (1857) | *Agrocybe aegerita* | 16.058 | 8.510e-3 |
| a.I2.m.C.parasitica.B.SSU.952 (2110) | *Cryphonectria parasitica* | 13.329 | 3.595e-2 |
| a.I2.m.K.lactis.A.OX1.i1 (2621) | *Kluyveromyces lactis* | 15.940 | 9.057e-3 |
| a.I2.m.M.polymorpha.A.SSU.911 (1640) | *Marchantia polymorpha* | 19.537 | 1.356e-3 |
| a.I2.m.P.anserina.A.ND5.i4 (2729) | *Podospora anserina* | 16.924 | 5.387e-3 |
| a.I2.m.P.hybrida.A.OX2.i1 (1456) | *Petunia x hybrid* | 20.950 | 6.430e-4 |
| a.I2.m.P.littoralis.B.LSU.575 (2440) | *Pylaiella littoralis* | 17.426 | 4.133e-3 |
| a.I2.m.P.sativum.B.S10.i1 (990) | *Pisum sativum* | 21.445 | 4.951e-4 |
| a.I2.m.S.cerevisiae.A.OX1.i1 (2520) | *Saccharomyces cerevisiae* | 16.273 | 7.597e-3 |
| a.I2.m.S.obliquus.B.LSU.2455 (625) | *Scenedesmus obliquus* | 14.832 | 1.626e-2 |
| a.I2.m.Z.mays.A.OX2.i1 (912) | *Zea mays* | 20.275 | 9.182e-04 |

RNA sequences and secondary structures were taken from [12]. Pseudoknots found in the a.I2.m.S.cerevisiae.A.OX1.i1 sequence (at positions 44-47:264-267, 50-55:312-317, 58-59:172-173, and 105-113:330-338) were removed before calculating the CWT. The CWT was calculated using Eqn. (2). The folding rate (in units of $sec^{-1}$) was estimated using the following equation (see the legend of Figure 1): Folding rate = exp($-5.2798\times10^{-1}$ CWT + 3.7118).

## Supplementary material

**Table S1.** Sequences and secondary structures of RNA sequences

| Name of RNA | Sequence | Secondary structure |
|---|---|---|
| *B. Subtilis* P RNA | GCGAGAAACCCAAAUUUUGGUAGGGGAACCUU<br>CUUAACGGAAUUCAACGGAGGGAAGGACAGAA<br>UGCUUUCUGUAGAUAGAUGAUUGCCGCCUGAG<br>UACGAGGUGAUGAGCCGUUUGCAGUACGAUGG<br>AACAAAACAUGGCUUAACGAACGUUAGACCACU<br>UACAUUUGGGAUCCUAACGUUCGGGUAAUCGC<br>UGCAGAUCUUGAAUCUGUAGAGGAAAGUCCAU<br>GCUCGCACGGUGCUGAGAUGCCCGUAGUGU | (((((...(((...........)))...(((((((................)))))))<br>((((((....))))))........(((((((.((...((((..(((.....)))..<br>.....))))...))................((((((((((...................<br>)))))))))))))))))).((((((((.....)))))))).............((<br>(((((((.((......))))))).....  |
| *B. Subtilis* specificity-domain | CUGCCUAGCGAAGUCAUAAGCUAGGGCAGUCUU<br>UAGAGGCUGACGGCAGGAAAAAAGCCUACGUC<br>UUCGGAUAUGGCUGAGUAUCCUUGAAAGUGCC<br>ACAGUGACGAAGUCUCACUAGAAAUGGUGAGA<br>GUGGAACCCGGUAAACCCCUC | (((((((((..........))))))(((((((....)))))).)((.((((...<br>..(((.(.(((....))).))))......))))....((..............(((<br>(((((....))))))))..............))))))) |
| *B. Subtilis* C-domain | GUUCUUAACGUUCGGGUAAUCGCUGCAGAUCU<br>UGAAUCUGUAGAGGAAAGUCCAUGCUCGCACG<br>GUGCUGAGAUGCCCGUAGUGUUCFFFGAGCGAG<br>AAACCCAAAUUUUGGUAGGGGAACCUUCUUAA<br>CGGAAUUCAACGGAGAGAAGGACAGAAUGCUU<br>UCUGUAGAUAGAUGAUUGCCGCCUGAGUACGA<br>GGUGAUGAGCCGUUUGCAGUACGAUGGAACAA<br>AACAUGGCUUACAGAACGUUAGACCACUU | ...((((((((((((((((((.((((((((.....))))))))...........<br>.(((((((((.((......))))))).....((...)))))))...(((........<br>...)))...(((((((................)))))))((((((....))))))).<br>.......))))))).((...((((..(((.....)))......))))...)).....<br>............)))))))))))...... |
| *Azoarcus* group I intron | GAGCCUUGCGCCGGGAAACCACGCAAGGGAUGG<br>UGUCAAAUUCGGCGAAACCUAAGCGCCCGCCCG<br>GGCGUAUGGCAACGCCGAGCCAAGCUUCGGCGC<br>CUGCGCCGAUGAAGGUGUAGAGACUAGACGGC<br>ACCCACCUAAGGCAAACGCUAUGGUGAAGGCAU<br>AGUCCAGGGAGUGGCGAAAGUCACACAAACC | ...((((((((..((....)).))))))).............((((((...((...((<br>((((....))))))..))...))))))(((.....((((((....))))))....<br>))).....(((((.........((((...(((....)))..))))......))))).<br>..((..((((....))))......)) |
| Hairpin ribozyme, four-way junction | CCGACAGAGAAGUCAACCAGAGAAACACACUUG<br>CGGCCGCAAGUGGUAUAUUACCUGGUACGCGUG<br>GUACCUGACAGUCCUGUCGG | (((((((....(((((((((.......((((((((()))))))))).........<br>)))))((()))((()))))))....))))))) |
| *Tetrahymena* group I intron P5abc domain | CCGUUCAGUACCAAGUCUCAGGGGAAACUUUGA<br>GAUGGCCUUGCAAAGGGUAUGGUAAUAAGCUG<br>ACGGACA | (((.((((((((..(((((((((....)))))))))..(((.....)))....)<br>))....).))))))... |
| *Tetrahymena* group I intron P4-P6 domain | AAUUGCGGGAAAGGGGUCAACAGCCGUUCAGU<br>ACCAAGUCUCAGGGGAAACUUUGAGAUGGCCU<br>UGCAAAGGGUAUGGUAAUAAGCUGACGGACAU<br>GGUCCUAACCACGCAGCCAAGUCCUAAGUCAAC<br>AGAUCUUCUGUUGAUAUGGAUGCAGUUCA | ...((((((....((((((....(((.((((((((..(((((((((....))))<br>)))))..(((.....)))....)))....).)))))))...))))))...)).))<br>))((...((((...((((((((.....))))))))..))))...))... |
| *S. cereviseae* | GCGGAUUUAGCUCAGUUGGGAGAGCGCCAGAC | ((((((...((((........)))).(((((.......))))).....(((((... |
| phenylalanil-tRNA | UGAAGAUCUGGAGGUCCUGUGUUCGAUCCACA GAAUUCGCACCA | ....))))).))))))).... |
| Hairpin ribozyme, two-way junction | AAAUAGAGAAGCGAACCAGAGAAACACACGCCA AAAUAUAUUUGGCGUGGUACAUACCUGGUACC CCCUCGCAUCCUAUUU | ((((((((.((((((((((((((.((((((((((.....))))))))))). ).))))))))))......))))))))))))) |
| *Tetrahymena* group I intron | AAAUAGCAAUAUUUACCUUUGGAGGGAAAAGU UAUCAGGCAUGCACCUGGUAGCUAGUCUUUAA ACCAAUAGAUUGCAUCGGUUUAAAAGGCAAGA CCGUCAAAUUGCGGGAAAGGGGUCAACAGCCGU UCAGUACCAAGUCUCAGGGGAAACUUUGAGAU GGCCUUGCAAAGGGUAUGGUAAUAAGCUGACG GACAUGGUCCUAACCACGCAGCCAAGUCCUAAG UCAACAGAUCUUCUGUUGAUAUGGAUGCAGUU CACAGACUAAAUGUCGGUCGGGGAAGAUGUAU UCUUCUCAUAAGAUAUAGUCGGACCUCUCCUUA AUGGGAGCUAGCGGAUGAAGUGAUGCAACACU GGAGCCGCUGGGAACUAAUUUGUAUGCGAAAG UAUAUUGAUUAGUUUUGGAGUACUCG | (((..............)))..........((((((((((......)))))))))).( ((((((((((.((.......)).)))))).)))))..............((((((.. ..((((((....(((.((((((((..(((((((((....)))))))))..((( .....)))....)))....).)))))))...))))))...)).))))((...((( (...((((((((.....))))))))..))))...))...(.(((((.......... .(((((((.....)))))))........))))))((...(((((....)))))(( (((((....((((......))))....))))))).((((((((.((((((.... ))))))).)))))))).........)). |

Sequences and secondary structures were taken from [13-18].

## MATLAB code for calculating the CWT metric

```
function [ cwt ] = CWT( in_args )
%This computer code calculates the contact-waiting-time (CWT) metric
%described in Nkwanta and Ndifon, FEBS Letters, 2009. It takes as input
%a MATLAB cell object containing either an RNA sequence and its
%pseudoknot-free secondary structure in the dot-brackets format, or a
% BPSEQ file containing both the sequence and secondary structure.
%Usage:
%   CWT({'AAAACCCUUU','((((...))))'})
%   CWT({",",'inputfile.bpseq'})

%the input sequence and structure
seq=in_args{1};
struc=in_args{2};
if(length(in_args)>2)
    fid = fopen(in_args{3});
    data = textscan(fid,'%s%s%s','Whitespace',' ');
    fclose(fid)
    temp=data{2}; seq=";
    temp2=data{3}; struc=";
    for i=1:length(temp);
        seq=cat(2,seq,temp{i});
        if(str2num(temp2{i})==0) struc=cat(2,struc,'.');
        else if(str2num(temp2{i})>i) struc=cat(2,struc,'(');
            else struc=cat(2,struc,')');
            end
        end
    end
end
seq=upper(seq);
cwt=0;
L=length(struc);
for i=1:L
    if(struc(i)=='(')
        f=0;
        j=i+1;
```

```
        while(f>-1 && j<=L)
          if(struc(j)=='(') f=f+1;
          else if(struc(j)==')') f=f-1; end
          end
          j=j+1;
        end
        if(j < L+1)
        pair=[seq(i) seq(j-1)];
        if(i<L && (struc(i+1)=='.' && struc(j-2)=='.'))
           cwt=cwt+getTime(i,j-1,1,pair);
        else
           cwt=cwt+getTime(i,j-1,0,pair);
        end
        end
    end
end
end

function time=getTime(left,right,nucl,pair)
%'pair' denotes a base pair (e.g., CG); 'left' the position of the base
% closest to the 5' end of seq; 'right' the other base; 'nucl'
%indicates whether or not the base pair is helix-nucleating
    time=0;
    R=1.9872e-3;
    T=273+37;
    if(sum(ismember(['CG';'GC'],pair,'rows'))>0); time=-3/(R*T);
    else if(sum(ismember(['AU';'UA'],pair,'rows'))>0) time=-2/(R*T);
        else if(sum(ismember(['GU';'UG'],pair,'rows'))>0) time=-1/(R*T);
            else if(pair(1)==pair(2)) time=1/(R*T);
                else time=0;
                end
            end
        end
    end
    if(nucl) time=time/log(right-left); end
    time=exp(time);
end
```